\documentclass[twocolumn,prb,showpacs,floatfix]{revtex4}
\usepackage{graphicx}
\usepackage{bm}
\usepackage{color}
\usepackage{amsmath}
\usepackage{amsfonts}
\usepackage{amssymb}

\begin{document}

\title{\textbf{Two-magnon excitations observed by neutron scattering in the\\
two-dimensional spin-5/2 Heisenberg antiferromagnet Rb$_2$MnF$_4$}}
\author{ T. Huberman, R. Coldea, R. A. Cowley}
\affiliation{Oxford Physics, Clarendon Laboratory, Parks Road, Oxford OX1 3PU, UK}
\author{D. A. Tennant}
\affiliation{School of Physics and Astronomy, University of St
Andrews, North Haugh, St Andrews, FIFE KY16 9SS, Scotland, UK}
\author{R. L. Leheny}
\affiliation{Department of Physics and Astronomy, Johns Hopkins University, Baltimore,
Maryland 21218, USA}
\author{R. J. Christianson}
\affiliation{Department of Physics $\&$ DEAS, Harvard University, Cambridge,
Massachusetts 02138, USA}
\author{C. D. Frost}
\affiliation{Rutherford Appleton Laboratory, Chilton, Didcot, Oxon OX11 0QX, UK}
\date{\today }
 % here go pacs numbers, classical and
%quantum spin models, dynamics}
% 75.40.Gb Dynamic properties (dynamic susceptibility, spin waves, spin diffusion, dynamic scaling, etc.)
%75.30.Ds spin waves
%75.10.Jm Quantized spin models
%75.50.Ee Antiferromagnets

\begin{abstract}
The low-temperature magnetic excitations of the two-dimensional
spin-5/2 square-lattice Heisenberg antiferromagnet Rb$_2$MnF$_4$
have been probed using pulsed inelastic neutron scattering. In
addition to dominant sharp peaks identified with one-magnon
excitations, a relatively weak continuum scattering is also
observed at higher energies. This is attributed to neutron
scattering by pairs of magnons and the observed intensities are
consistent with predictions of spin wave theory.
\end{abstract}

\maketitle

\section{Introduction}

Rb$_2$MnF$_4$ is a near-ideal two-dimensional spin-5/2 Heisenberg
antiferromagnet on a square lattice (2DHASL) and has been
extensively studied
experimentally\cite{Birgeneau1,Birgeneau2,Leheny1,cowley1} to test
theoretical predictions for this canonical 2D quantum Hamiltonian.
Neutron scattering experiments observe sharp one-magnon
excitations at low temperatures which could be well described by
linear spin wave theory.\cite{cowley1} However, spin wave theory
predicts that neutrons can also scatter from pairs of magnons
leading to a broad scattering continuum at higher energies. Such
processes are longitudinally polarized with respect to the ordered
spin direction and correspond physically to the simultaneous
creation of two magnons with opposite spin $S_z=\pm 1$ (total
$\Delta S_z=0$ process). The overall scattering weight of the
two-magnon continuum relative to one-magnon processes is related
to the relative strength of the zero-point longitudinal quantum
fluctuations in the ground state, which reduce the amount of
ordered spin moment by $\Delta S$ compared to the full spin value
$S$. Very sensitive experiments are required to search for
two-magnon continuum scattering since for spin-5/2 the two-magnon
scattering intensity integrated over energy and wavevector is
expected to be only of the order 6\% of the integrated one-magnon
intensity.

Another probe of two-magnon processes is light or Raman scattering
and experiments on several 2DHAFSL systems have been made.
Measurements in the $S$=1 system
K$_{2}$NiF$_{4}$~\cite{Fleury1,Chinn1} were found to be in
excellent agreement with calculations based on interacting spin
wave theory. Similar experiments on the cuprates ($S$=1/2) have,
however, shown a clear discrepancy between the observed lineshape
and calculations based on spin wave theory for pure
2DHAFSL.\cite{Lyons1} The discrepancy is believed to arise from
the presence of the four-spin cyclic exchange
terms,\cite{Sugai1,Roger1} which are also used to explain the
observed spin wave dispersion along the antiferromagnetic zone
boundary in La$_{2}$CuO$_{4}$ ref.\ \onlinecite{Coldea1}.

Neutron scattering (unlike Raman scattering which has inherent
momentum constraints) can in principle access the full wavevector
and energy dependence of the two-magnon scattering. Raman and
neutron scattering offer complementary information in that neutron
intensities are related to two-operator correlation functions,
whereas Raman is related to four-operator terms.

So far, two-magnon continuum scattering has been observed in the
3D material CoF$_2$~\cite{cowley2,Holden1}, but  little has been
done to quantitatively measure the two-magnon scattering using
neutrons in 2DHAFSL systems, the only other experiments we are
aware of are on the spin-1/2 material Cu(DCOO)$_2$.4D$_2$O
(CFTD).\cite{Christensen1} The large quantum corrections in
$S$=1/2 systems make it important to measure the two-magnon
scattering in Rb$_{2}$MnF$_{4}$ with $S=5/2$ where quantum
fluctuations are expected to be smaller and to test the extent to
which spin wave theory can describe the results.

%Two-magnon neutron scattering has a different matrix element
%compared to light scattering and the full wavevector dependence of
%the scattering also becomes accessible in addition to the energy
%dependence.
%
%so new information and slso Neutron scattering from two magnons
%has a different scattering matrix element
%
%The interaction of neutron and magnons is much better understood
%
%New information can be obtained from neutron scattering tow-magnon
%scattering from tfrom magnons has a different matrix element and
%can also probe the wavevector-dependence so

%provide a  because here the low temperature properties are well
%understood and can be described by linear spin wave theory.

%apart from on CFTD~\cite{Christensen1}, which has $S$=1/2. The
%large quantum corrections in $S$=1/2 systems make it important to
%measure the two-magnon scattering in Rb$_{2}$MnF$_{4}$ because the
%low temperature properties are well understood and can be
%explained by linear spin wave theory.

%The dispersion relation in Rb$_{2}$MnF$_{4}$ has been measured
%using triple-axis neutron scattering\cite{cowley1} which
%determined the exchange $J =0.6544\pm 0.0014$ meV and anisotropy
%$\Delta =0.0048\pm 0.0010$.
The purpose of this paper is threefold. Firstly we wished to
measure the spin waves with the MAPS spectrometer at ISIS to
investigate whether high quality measurements of the dispersion
relation could be made with a neutron time-of-flight spectrometer
as compared with earlier triple-axis measurements.\cite{cowley1}
Secondly we wished to study the spin wave energy along the
antiferromagnetic zone boundary to search for evidence of
four-spin interactions, second neighbour exchange interactions or
quantum corrections to linear spin wave theory. Thirdly we wished
to look for the two-magnon scattering continuum predicted by spin
wave theory but lacking experimental evidence.

%The only other neutron scattering experiments we are aware of,
%which characterize the two-magnon scattering for the 2DHAFSL, are
%for the spin-1/2 material Cu(DCOO)$_2$.4D$_2$O
%(CFTD)\cite{Christensen1}. The large quantum corrections in
%$S$=1/2 systems make it important to measure the two-magnon
%scattering in Rb$_{2}$MnF$_{4}$ with $S=5/2$ because here the low
%temperature properties are well understood and can be described by
%linear spin wave theory.

%Quantum fluctuations are expected to be smaller in
%Rb$_{2}$MnF$_{4}$ with spin-5/2 and it is clearly of interest to
%study the two-magnon scattering in this material and see if it can
%be described by spin-wave theory.

The rest of the paper is organized as follows. In Sec.\
\ref{Rb2MnF4_correlations1} we review the spin wave theory
predictions for one- and two-magnon scattering. Details of the
experiment are given in Sec.\ \ref{Experimental_Details_Ch1},
followed in Sec.\ \ref{Dispersion_relation_Rb2MnF4} by results for
the main dispersion relation with particular attention to
wavevectors on the antiferromagnetic zone boundary. Sec.\
\ref{TwoMagnon} reports on the observation of two-magnon
scattering. The results are summarized in a final Sec\
\ref{Conclusions}.

\section{Dynamical correlations}

\label{Rb2MnF4_correlations1}

%Rb$_{2}$MnF$_{4}$ has been proposed as a near-ideal example of a
%two-dimensional spin-5/2 Heisenberg antiferromagnet on a square
%lattice (2DHAFSL). This model is not expected to have long range
%order at finite temperatures, but in many real systems, such as
%K$_{2}$NiF$_{4}$($S$ =1), there are additional subleading terms in
%the Hamiltonian which break the continuous symmetry and stabilize
%long-range order at low temperatures and those phases can be well
%described by spin wave theory.

Rb$_{2}$MnF$_{4}$ crystallizes in the tetragonal K$_{2}$NiF$_{4}$
structure with space group \emph{I}4/\emph{mmm} and lattice
parameters $a$ = $b$ = 4.215\AA, $c$ = 13.77\AA. The magnetic ions
are Mn$^{2+}$ with a spin-only moment of $S=5/2$, arranged in a
square lattice in the basal plane, with antiferromagnetic
superexchange couplings between nearest neighbours mediated by
intervening F$^{-}$ ions. The inter-plane coupling is a factor of
$10^{-4}$ of the intra-plane coupling~\cite{Wijn}. This almost
perfect two-dimensionality arises because MnF$_{2}$ magnetic
layers have a large separation along the $c$-axis filled by two
non-magnetic RbF sheets, and furthermore the magnetic couplings
along $c$ are frustrated because each Mn$^{2+}$ ion is equidistant
to four antiferromagnetically-coupled spins in the layers below
and above, further weakening the effect of the inter-plane
couplings.

Antiferromagnetic order with moments along the $c$-axis occurs
below $T_N=38.4$\ K. Earlier studies\cite{Birgeneau2} proposed
that the observed ordering at finite temperature can be well
accounted for by a small anisotropy $\delta_z=0.0048(10)$ in the
Hamiltonian, ultimately originating from dipolar interactions,
i.e.
\begin{equation} \hat{H}=J\sum_{\langle ii'\rangle}[S_{i}^{x}S_{i'}^{x}+S_{i}^{y}S_{i'}^{y}+
(1+\delta_z )S_{i}^{z}S_{i'}^{z}], \label{Hamiltonian}
\end{equation}
where $J=0.6544(14)$ meV [ref. \onlinecite{cowley1}] is the
exchange energy for nearest-neighbour spins on the square lattice
and $\langle ii' \rangle$ indicates that each interacting spin
pair is counted once in the summation. ($x$,$y$,$z$) are along the
crystallographic ($a$,$b$,$c$) axes. For this Hamiltonian the
magnon dispersion relation in linear spin wave theory
is\cite{cowley1}
\begin{equation}
\omega_{\bm{Q}}=4JS\left[\left(1+\delta_z
\right)^{2}-\gamma_{\bm{Q}}^{2}\right]^{\frac{1}{2}}
\label{eqn:1_S_dispersion}
\end{equation}
where $\gamma_{\bm{Q}} =\cos \pi (Q_{k}+Q_{h})\cos \pi
(-Q_{k}+Q_{h})$ and ($Q_h,Q_k,Q_l$) are components of the crystal
momentum ${\bm Q}$ given in rlu units of ($2\pi /a,2\pi /b,2\pi
/c)$. Often a multiplicative factor, $Z_c=(1+0.157/2S)$, is
included in the dispersion relation to account for corrections to
lowest order spin wave theory. For $S=5/2$, $Z_c=1.0314$, and we
neglect this correction as it can be readily incorporated into the
exchange constant.

Neutron scattering measures the  dynamical correlation functions
given by
\begin{eqnarray}  \label{eqn:correlations_antiferromnagnet}
&& S^{\alpha \alpha }(\bm{Q},\omega ) =  \notag \\
&&\frac{1}{2\pi \hbar N} \int_{-\infty }^{\infty }dte^{-i\omega
t}\sum_{ii^{\prime }}e^{i\bm{Q}.(\bm{r}_{i}-\bm{r}_{i^{\prime
}})}\langle S_{i^{\prime }}^{\alpha }(0)S_{i}^{\alpha }(t)\rangle  \notag \\
\end{eqnarray}
where $N$ is the total number of spins and the sum runs over all
sites $i$ and $i^{\prime}$ in the lattice. One-magnon events occur
in the spin correlations {\em transverse} to the ordered spin
direction $z$. In the non-interacting spin wave approximation the
transverse correlations at $T=0$\ K are given by\cite{Heilmann1}
\begin{eqnarray}
S^{xx}(\bm{Q},\omega )&=&S^{yy}(\bm{Q},\omega )\nonumber
\\
& = & \frac{1}{2}(S-\Delta S)(u_{\bm{Q}}+v_{\bm{Q}})^{2} \delta
(\hbar \omega -\hbar \omega _{\bm{Q}})
\label{eqn:one_magnon_corellations}
\end{eqnarray}
where $u_{\bm{Q}}=\cosh \theta$, $v_{\bm{Q}}=\sinh \theta$, $\tanh
2\theta=-\frac{\gamma_{\bm{Q}}}{(1+\delta_z)}$. Here $\Delta
S=S-\langle S^z \rangle$ is the spin reduction due to zero-point
fluctuations calculated as $(1/N)\sum_{\bm Q} v^2_{\bm{Q}}$ where
the sum extends over $\bm{Q}^{\prime}$s in the full Brillouin
zone. $\Delta S = 0.197$ for the isotropic Heisenberg model and
0.167 for the anisotropy $\delta_z$ appropriate for Rb$_2$MnF$_4$.

The finite spin reduction allows for the presence of {\em
longitudinal} fluctuations, which can be described in terms of
two-magnon scattering events. In the non-interacting spin wave
approximation the longitudinal correlations at $T=0$\ K
are\cite{Heilmann1}
\begin{eqnarray}
& & S^{zz}(\bm{Q},\omega )_{\mathrm{inelastic}}=\frac{1}{N}
\sum_{\bm{Q}_1,\bm{Q}_2}
f \left(\bm{Q}_1,\bm{Q}_2 \right) \times \notag \\
& & ~~~~\delta({\bm Q}- {\bm Q}_1+{\bm Q}_2) ~ \delta (\hbar
\omega -\hbar \omega _{\bm{Q} _{1}}-\hbar \omega _{\bm{Q}
_{2}})\label{eqn:Szz_Rb2MnF4_lowT}
\end{eqnarray}
where $f({\bm Q}_1,{\bm Q}_2)= \frac{1}{2}\left(
u_{\bm{Q}_{1}}v_{\bm{Q}_{2}}-u_{\bm{Q}_{2}}v_{
\bm{Q}_{1}}\right)^2$ is the structure factor for creating two
magnons at wavevectors ${\bm Q}_1$ and ${\bm Q}_2$. In the
summation, one of the two magnons (say $\bm{Q}_1$) is restricted
to the first Brillouin zone. The above equation gives the
inelastic part of the longitudinal correlations, the elastic part
is simply the Bragg peak contribution $(S-\Delta
S)^2\delta(\hbar\omega)~\delta(\bm{Q}-\bm{Q}_{\mathrm{AF}} -
\bm{\tau})$, where $\bm{Q}_{\mathrm{AF}}$=(0.5,0.5) is the
antiferromagnetic ordering wavevector and $\bm{\tau}$ is a vector
of the reciprocal lattice.

An understanding of how the scattering cross-section is
distributed between the elastic, one and two-magnon channels can
be obtained by comparing the integrated scattered intensities. The
total intensity integrated over energy and averaged over the
Brillouin zone is given by the well known sum rule $\int
S(\bm{Q},\omega) d \bm{Q} d(\hbar\omega)=S(S+1)$. Similar
expressions can be derived for the individual scattering
components, by integrating over the expressions for the elastic,
one-magnon eq.\ (\ref{eqn:one_magnon_corellations}) and two-magnon
eq.\ (\ref{eqn:Szz_Rb2MnF4_lowT}) components. The results are
summarised in Table\ \ref{table:int_intensity}.
\begin{table}[h!]
\begin{tabular}{|c|c|}
\hline component & integrated intensity \\ \hline
$S(\bm{Q},\omega)$ & $S(S+1)$= 8.75 \\
$S^{zz}(\bm{Q},\omega)_{\mathrm{elastic}}$ & $(S-\Delta S)^2$ = 5.443 \\
$S^{xx}(\bm{Q},\omega)+S^{yy}(\bm{Q},\omega)$ & $(S-\Delta S)(2
\Delta S + 1)$ = 3.112 \\
$S^{zz}(\bm{Q},\omega)_{\mathrm{inelastic}}$ & $\Delta S (\Delta S
+1)$ = 0.195 \\ \hline
\end{tabular}
\centering \caption{Total sum rules for the different components
of the scattering, evaluated for $S=5/2$ and $\Delta S = 0.167$.}
\label{table:int_intensity}
\end{table}

\begin{figure*}[ht!]
\caption{(Color online) (a) One-magnon dispersion surface as a
function of two-dimensional wavevector ($Q_h,Q_k$) and energy
$\protect\omega$ (color shading is intensity in neutron
scattering). Dashed lines in the basal plane and at maximum energy
$\hbar\omega=4JS$ mark the antiferromagnetic zone boundaries. The
basal plane also shows constant-energy contours (solid lines). (b)
Constant energy maps of the magnetic scattering at $\hbar\omega=$
1, 3.5 and 6\ meV obtained by taking slices from the 3D
($Q_h,Q_k,\protect\omega$) neutron data. (c) Energy scan at
constant wavevector ($-0.5,0$) along the direction shown by the
vertical rectangular column in (b) (cross-section of column
$\Delta Q_h \times \Delta Q_k$ indicates the wavevector region
over which intensity points were averaged).}
\label{fig:zone_boundary}
\begin{center}
\includegraphics*[width=17.5cm,bbllx=9,bblly=584,bburx=592, bbury=828]
 {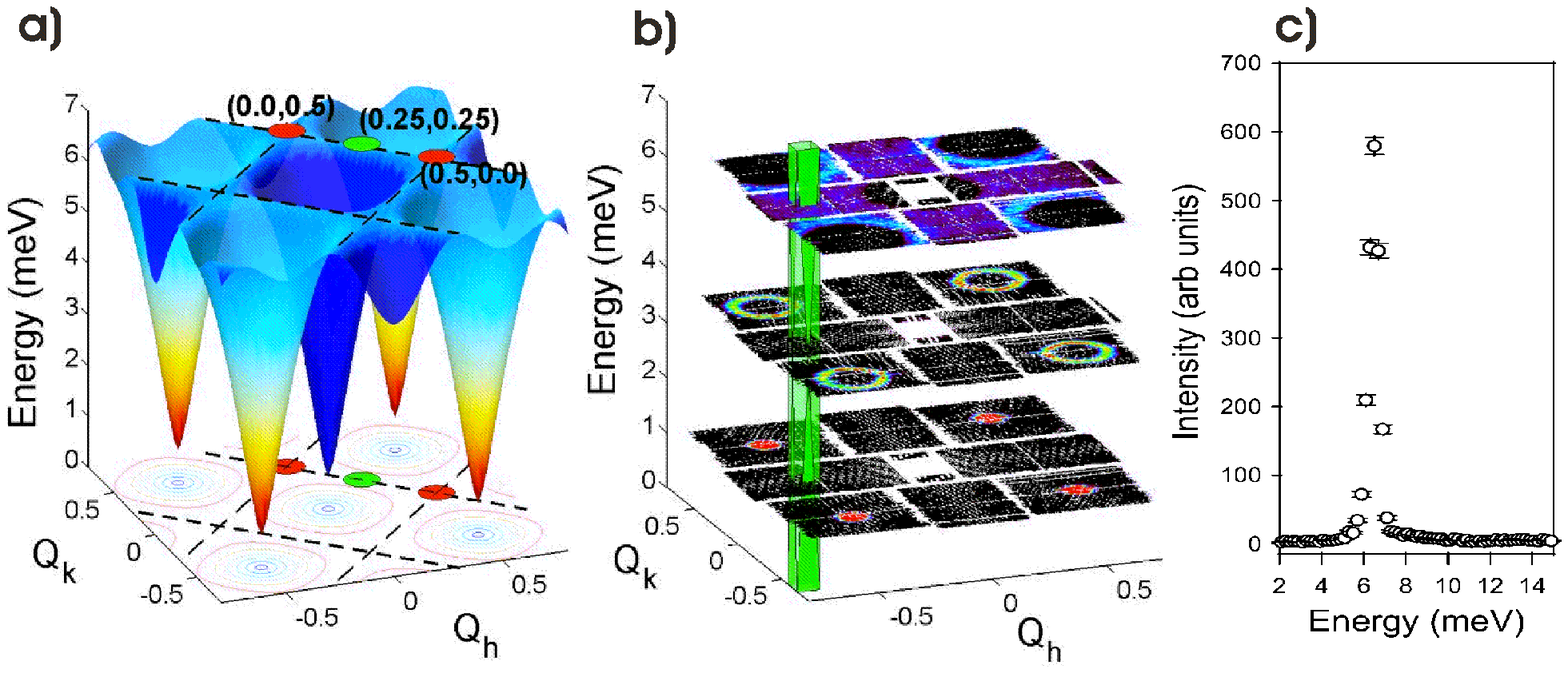}
\par
%\caption{Graph a) shows the one-magnon correlation functions at
%$T=0$. The shape of the curve is given by the spin wave dispersion
%relation and the colour marks the intensity. The dashed lines mark
%the antiferromagnetic zone boundary. Also marked are the points
%(0.5,0), (0,0.5) and (0.25,0.25), which are referred to
%specifically in the discussion regarding measurement of the change
%in energy a long the zone boundary in section~\ref{SelfEnergy}. In
%b), some constant energy slices through the data are shown and
%these can be compared with panel a). The green column in b) is an
%example of the volume of space sampled when taking a
%constant-$\bm{Q}$ cut through the data. The resulting
%lineshape is displayed in c)}
\end{center}
\end{figure*}

The inelastic neutron scattering intensity from one and two-magnon
excitations is proportional to
\begin{equation}
(2-p_z) S^{xx}(\bm{Q},\omega )+ p_z S^{zz}(\bm{Q},\omega)_{\mathrm
{inelastic}} \label{eqn:scattering_x_section}
\end{equation}
%_{\mathrm{inelastic}}
where $p_z=1-\hat{Q}^2_z$ is the polarization factor for
longitudinal scattering, $\hat{Q}_z$ is the directional cosine of
the wavevector ${\bm Q}$ with respect to the $c$-axis. The
proportionality factor between eq.\
(\ref{eqn:scattering_x_section}) and neutron scattering intensity
includes the magnetic form factor squared $F(\bm{Q})^{2}$ of
Mn$^{2+}$ ions.

%\begin{eqnarray}
%&&\frac{d\sigma }{d\Omega dE_{f}}\propto
%F(\bm{Q})^{2}\sum_{\alpha }(1- \hat{Q}_{\alpha
%}^{2}){S}^{\alpha \alpha }(\bm{Q},\omega )  \notag
%\label{eqn:scattering_x_section} \\
%&=&F(\bm{Q})^{2}[p_T S^{xx}(\bm{Q},\omega )+p_LS^{zz}(\bm{Q},\omega )_{\mathrm{inelastic}}]  \notag \\
%&&
%\end{eqnarray}
%Where the transverse and longitudinal polarization factors are
%given by $p_T=1+\hat{Q}_{z}^{2}$ and $p_L=1-\hat{Q}_{z}^{2}$.

\section{Experimental Details}

\label{Experimental_Details_Ch1}

The magnetic scattering was measured from a 13.4g single crystal
of Rb$_{2}$MnF$_{4}$ using the MAPS spectrometer at the ISIS
pulsed neutron source in the UK. The sample was enclosed in an
aluminium can containing helium exchange gas and measurements
reported here were made at the base temperature of 9.5~K. MAPS is
a direct geometry time-of-flight instrument, equipped with a 16\
m$^{2}$ array of position sensitive detectors, divided into nearly
$4\times10^4$ separate detector elements. This allows collection
of the inelastic scattering intensity in a highly pixelated 3D
volume in the 4D ($Q_h,Q_k,Q_l,\omega$) space, from which one can
extract the intensity plot in a certain plane or along a certain
direction as illustrated in Fig.~\ref{fig:zone_boundary}b) and c).
We describe the data in terms of the two in-plane wavevectors
($Q_{h},Q_{k}$) and energy $\hbar\omega$ as the magnetism is
two-dimensional and the inter-layer component $Q_{l}$ only enters
the magnetic scattering through the magnetic form factor
$F(\bm{Q})$ and the polarization factor $p_z$ with respect to the
ordered spin direction, see eq.\ (\ref{eqn:scattering_x_section}).

%Constant $\bm{Q}$ cuts can be obtained from the data by
%averaging over all the scattering points within a small region of
%space $(Q_{h}\pm \Delta Q_{h},Q_{k}\pm \Delta Q_{k})$ and assuming
%the data is representative of the mean value for $\bm{Q}_{M}$
%in the cut region. Using this notation, a constant-$\bm{Q}$
%cut will be said to have dimensions $\Delta Q_{h}\times \Delta
%Q_{k}$. An example of the volume of $\bm{Q}$-$\omega $ sampled
%by a cut is shown in figure~\ref{fig:zone_boundary}b) with the
%resulting cut lineshape shown in figure~\ref{fig:zone_boundary}c).

\begin{figure}[ht!]
\begin{center}
\includegraphics*[width=8cm,bbllx=3,bblly=158,bburx=592, bbury=747]{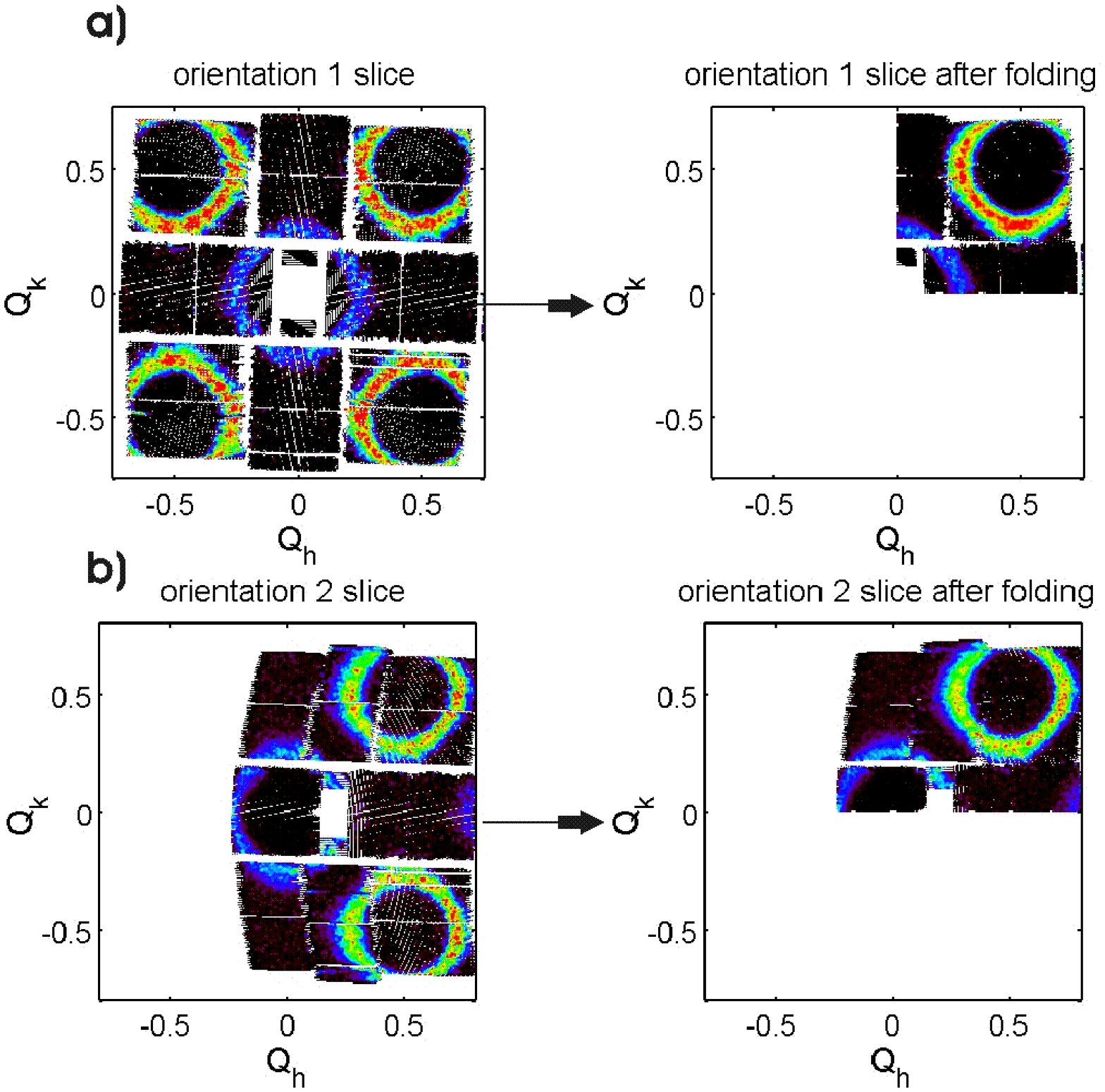}
\end{center}
\caption{(Color online) Folding of data along symmetry-equivalent
directions: (a) four-fold symmetry in the $\bm{c} \parallel
\bm{k}_i$ setup (b) two-fold symmetry in the rotated
configuration. Data corresponds to intensity at an energy $ 5.5\pm
0.5$ meV.} \label{fig:folding_slices}
\end{figure}

An incident neutron energy of $24.92$ meV was selected to map the
inelastic scattering over the whole dynamic range of one- and
two-magnon scattering processes which extended up to 13 meV. A
Fermi chopper spinning at 300\ Hz gave an energy resolution of
0.75$\pm$ 0.01meV (FWHM) on the elastic line. Measurements were
made with the two-dimensional magnetic layers arranged
perpendicular to the incident neutron beam (orientation 1, $\bm{c}
\parallel \bm{k}_i$, $\bm{a}$ horizontal) and tilted by an angle $\psi=45^{\circ}$
with respect to the incident beam direction (orientation 2,
$\hat{(\bm{c},\bm{k}_i)}=\psi$) to collect complementary data with
different longitudinal polarization $p_z$ at the same
two-dimensional wavevector and energy ($Q_h,Q_k,\omega$). Typical
counting times for one crystal orientations were 20 hours at an
average proton current of 170 $\mu$A. To increase statistics the
data was folded along symmetry-equivalent axes as illustrated in
Fig.\ \ref{fig:folding_slices}, four-fold in the $\bm{c}$-axis
along $\bm{k}_i$ setup, and two-fold in the rotated configuration.
The low angle detector bank $2\theta<30^{\circ}$ provided coverage
over most of the first Brillouin zone.
%MAPS also has a high angle detector bank which gives data for
%$Q_{h}\geq 1.8$. We have chosen not to use this part of the data
%because the data has a small range in $Q_{k}$, $0.2\leq Q_{k}\leq
%0.2$ and so does not cover such a large region in reciprocal
%space, as well as being in a region where both the one and
%two-magnon cross-sections have low intensities. These aspects mean
%that the overall quality of this data is not as good as the data
%from the low angle bank.

\section{Spin wave dispersion relation}

\label{Dispersion_relation_Rb2MnF4}

\begin{figure}[ht!]
\begin{center}
\includegraphics*[width=8cm,bbllx=13,bblly=34,bburx=583, bbury=819]{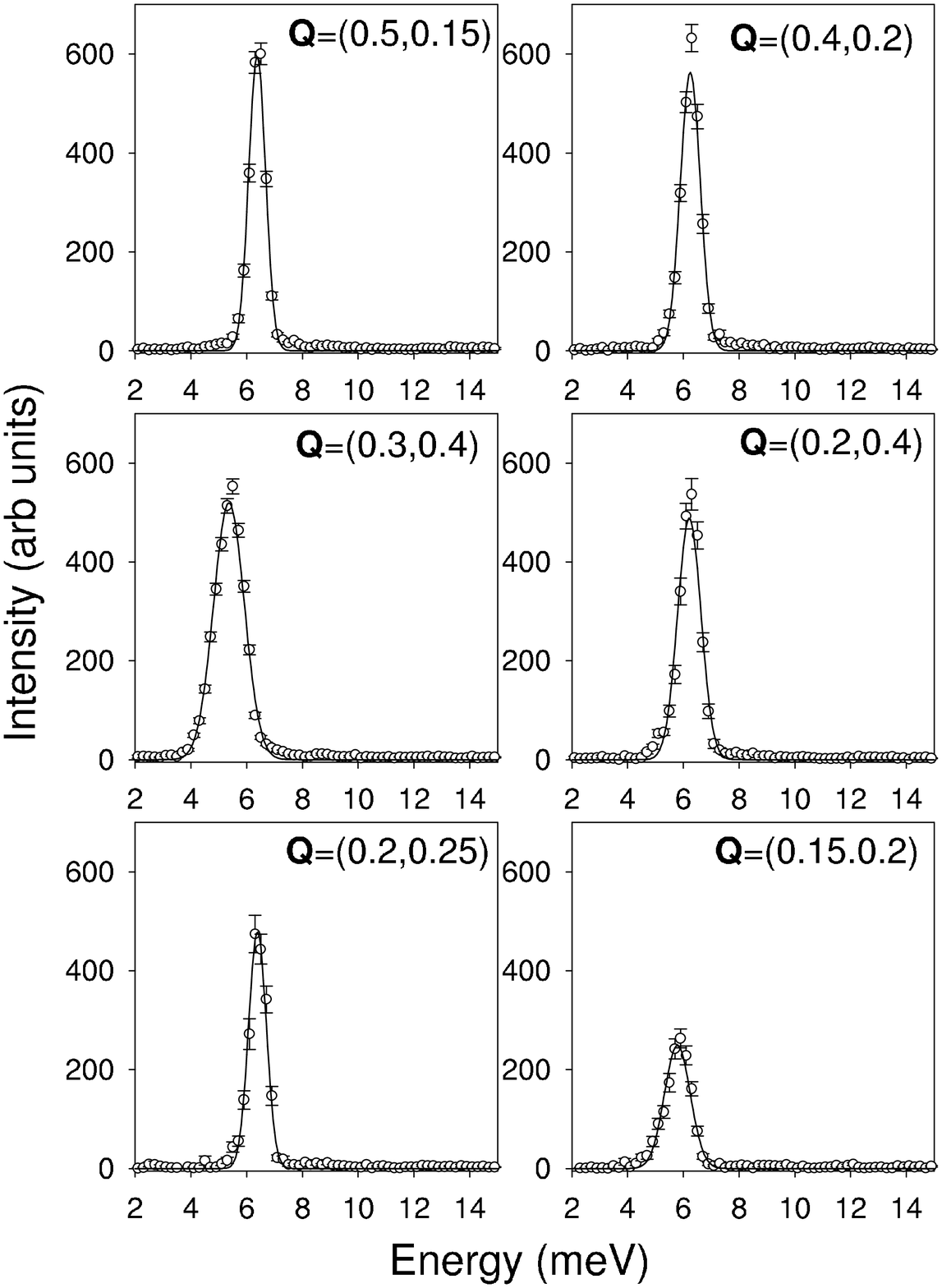}
\caption{Energy scans at constant ($Q_h,Q_k$) wavevector. Solid
lines are fits to Gaussian peaks.} \label{fig:basic_cuts}
\end{center}
\end{figure}

\begin{figure}[ht!]
\begin{center}
\includegraphics*[width=8cm,bbllx=14,bblly=445,bburx=585, bbury=822]{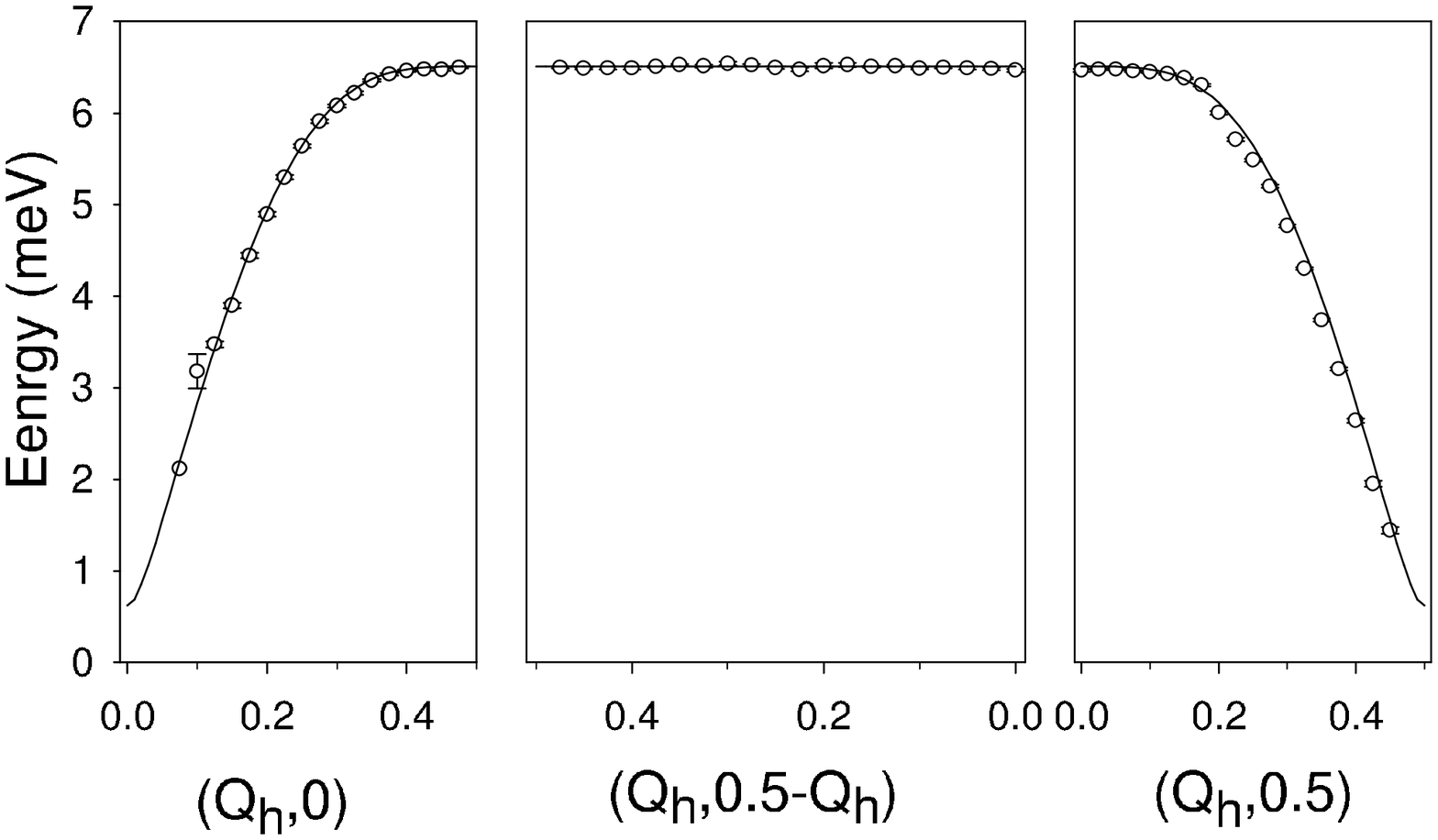}
\caption{Dispersion relation along symmetry directions in the
Brillouin zone (bold lines in Fig.\ \ref{fig:two_slices_magnonx}
b)). Solid line shows a fit to eq.\
(\protect\ref{eqn:1_S_dispersion}) with $J=0.648$ meV.}
\label{fig:basic_dispersion}
\end{center}
\end{figure}

The magnon dispersion relation was determined from fits to energy
scans at constant wavevector ($Q_h,Q_k$) and typical data is shown
in Fig.\ \ref{fig:basic_cuts}. The scattering is dominated by a
sharp one-magnon excitation and solid lines show fits to eq.\
(\ref{eqn:one_magnon_corellations}), where the delta function
$\delta (\hbar \omega -\hbar \omega _{\bm{Q}})$ is replaced by a
resolution broadened Gaussian peak.
%The peak centre is largely unaffected by the resolution broadening
%and was found by fitting to a Gaussian curve. The centre then
%gives the spin wave energy at the cut wave-vector. Using these
%results, the spin wave dispersion was determined and
The extracted spin wave dispersion along symmetry directions in
the Brillouin zone is shown in Fig.~\ref{fig:basic_dispersion}.
Data at the lowest energies is limited because of the difficulty
in resolving the one-magnon peak from the elastic incoherent
scattering. The solid line shows a fit to the dispersion relation
in eq.\ (\ref{eqn:1_S_dispersion}) with a fitted exchange
$J=0.648\pm 0.003$ meV and fixed anisotropy $\delta_z=0.0048$, in
agreement with the previous estimates of $J=0.6544 \pm 0.0014$ meV
obtained from triple-axis neutron measurements\cite{cowley1}.

%\section{Detailed Measurements of the Dispersion Relation}
%\label{SelfEnergy}

Linear spin wave theory, eq.\ (\ref{eqn:1_S_dispersion}), predicts
no dispersion along the antiferromagnetic zone boundary (central
panel in Fig.\ \ref{fig:basic_dispersion}) therefore an observed
dispersion along this direction can give information about
failures of the linear spin wave approximation or extra
interactions in the Hamiltonian beyond the nearest-neighbour
exchange. In the spin-1/2 organic material CFTD~\cite{Ronnow1},
the zone boundary energy was observed to decrease from (0.25,0.25)
to (0.5,0) by $6 \pm 1\%$, and this agreed with computational work
on the $S=1/2$ 2DHAFSL
using the Ising-limit expansion that predicts a dispersion of $%
7\%$~\cite{Singh2}. The zone-boundary dispersion was therefore
attributed to quantum corrections to linear spin wave theory. A
spin wave calculation extended to order $1/S^{2}$ predicts the
same sign but a much smaller magnitude ($2\%$) of this
dispersion\cite{Canali1} than the series results, suggesting that
higher order corrections would need to be considered to obtain a
fully satisfactory theory. In La$_{2}$CuO$_{4}$ the observed
zone-boundary dispersion had the opposite sign, increasing from
(0.25,0.25) to (0.5,0) by $13\%$ ~\cite{Coldea1}. This effect was
attributed to higher-order spin exchange terms in the Hamiltonian.
La$_2$CuO$_4$ is a Mott insulator and a more appropriate
description of the electronic states is in terms of a Hubbard
model at half-filling, characterized by a kinetic energy gain from
hopping, $t$, and potential energy cost, $U$, for two electrons
occupying the same site. At small $t$ electrons are localized, and
perturbative expansion in $t/U$ gives an effective Hamiltonian for
the spin degrees of freedom. The first term in this expansion is
the nearest-neighbour Heisenberg exchange, and the dominant next
order term is a cyclic exchange coupling 4 spins at the corners of
each square plaquette.
%To fourth order in
%$t/U$ there are cyclic exchange terms in addition to the
%Under certain conditions~\cite{Chao1}, the kinetic energy can be
%expanded perturbatively, in powers of $t^{m}/U^{m-1}$.
%At fourth order a cyclic exchange coupling 4 spins at the corders of each
%square plaquette occurs and this.
Such a ring exchange term corresponding to $U/t\sim6$ was used to
describe the high-energy magnon dispersion,\cite{Coldea1} Raman
scattering\cite{Sugai1} and infrared absorption
experiments\cite{Lorenzanna1} in La$_2$CuO$_4$.

\begin{figure}[ht!]
\begin{center}
\includegraphics*[width=8cm,bbllx=20,bblly=520,bburx=585, bbury=835]{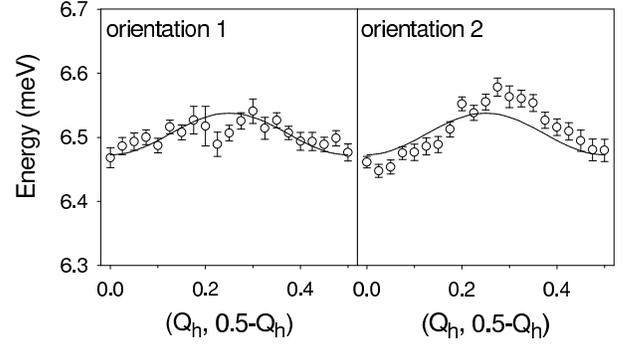}
\end{center}
\caption{Magnon energy for wavevectors along the antiferromagnetic
zone boundary obtained using narrow ($ 0.04\times 0.04$) cuts from
data collected under two different experimental setups. Solid line
is a fit to the spin wave dispersion relation including
next-nearest neighbour coupling, eq.\
(\ref{eqn:next_nearest_neighbour_dispersion}).}
\label{fig:zone_boundary_dispersion}
\end{figure}

\begin{figure}[ht!]
\begin{center}
\includegraphics*[width=8cm,bbllx=20,bblly=435,bburx=585, bbury=830]{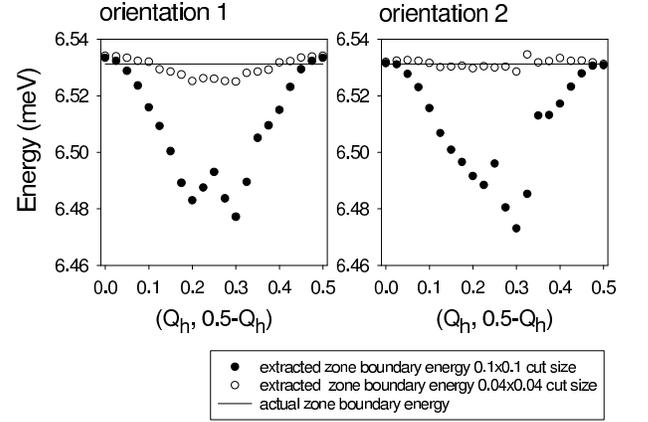}
\end{center}
\caption{Energy of one-magnon peak extracted from cuts through a
simulated data set for eqs.\
(\ref{eqn:1_S_dispersion},\ref{eqn:one_magnon_corellations}).
Using a large wavevector averaging range $\Delta Q_h \times \Delta
Q_k=0.1\times 0.1$ around the nominal ($Q_h,Q_k$) position
introduces an apparent dispersion of $0.9\%$ (filled circles),
whereas the smaller cut size $0.04\times 0.04$ (open symbols)
reduces this to less than 0.1\%.}
\label{fig:simulated_zone_boundary_dispersion}
\end{figure}

The spin wave energies along the antiferromagnetic zone-boundary
contour in Rb$_2$MnF$_4$ are shown in a blown-out scale in
Fig.~\ref{fig:zone_boundary_dispersion}. The dispersion along the
$Q_{h}+Q_{k}=0.5$ direction is very small, of order $1\pm0.5\%$
between (0.5,0) and (0.25,0.25). In extracting peak positions we
carefully considered the effects of wavevector averaging over a
box of finite size $\Delta Q_h \times \Delta Q_k$ around the
nominal ($Q_h,Q_k$) values. Since the dispersion surface has a
maximum on the zone boundary, the effect of a finite wavevector
averaging is to produce an apparently lower peak energy. The
effect is more pronounced around the (0.25,0.25) point since there
the dispersion surface has only a one-dimensional maximum, whereas
the corner point (0.5,0) is a local maximum along two directions
in the plane, see Fig.\ \ref{fig:zone_boundary}a). This effect is
illustrated in Fig.~\ref{fig:simulated_zone_boundary_dispersion}
by taking cuts of different sizes $\Delta Q_h \times \Delta Q_k$
over a simulated data set for the one-magnon scattering
cross-section, eq.\ (\ref{eqn:one_magnon_corellations}). A
wavevector box of size $0.04\times0.04$ was chosen as a balance
between a minimal apparent peak shift ($<0.1\%$) and sufficient
data pixels in the box for sufficient statistics to allow
quantitative analysis and the final results are shown in
Fig.~\ref{fig:zone_boundary_dispersion}. Data collected under two
sample orientations show a small dispersion with the higher energy
at (0.25,0.25). The magnitude of the dispersion is small,
$1\pm0.5\%$ of the zone boundary energy, a value close to the
limit of the experimental accuracy, which may explain why the two
data sets are not exactly overlapping.

%Plots of the spin wave energy, extracted from the neutron scattering data,
%along the zone boundary are shown in figure~\ref%
%{fig:zone_boundary_dispersion}. The results from both orientations show that
%there is a marked maximum in the centre of the zone boundary but that the
%magnitude of the effect depends on the orientation. Clearly both
%orientations of the crystal should give identical dispersion curves, but we
%have been unable to decide which of the results corresponds to the correct
%result. We therefore estimate from the measurements that the difference in
%frequencies is 0.07$\pm 0.04$ meV with the highest energy at the (0.25,0.25)
%zone boundary.

We note that quantum corrections to linear spin wave theory to
order $1/S^{2}$ would predict a dispersion of the same sign but
nearly an order of magnitude smaller, $0.005$ meV~\cite{Canali1}.
Earlier triple-axis measurements also observed a dispersion
similar to the one in the present experiments and Cowley \emph{et
al}.~\cite{cowley1} proposed that the origin was a next-nearest
neighbour antiferromagnetic exchange $J^{\prime}$ along the square
diagonals. The dispersion relation in this case becomes
\begin{equation}
\hbar\omega _{\bm{Q}}=4JS\left[\left(1+\delta_z
+\frac{J^{\prime}}{2J}\gamma^{\prime}_{
\bm{Q}}\right)^{2}-\gamma_{\bm{Q}}^{2}\right]^{1/2},
\label{eqn:next_nearest_neighbour_dispersion}
\end{equation}
where $\gamma_{\bm{Q}}=\cos \pi (Q_{k}+Q_{h})\cos \pi
(-Q_{k}+Q_{h})$, $\gamma^{\prime}_{\bm{Q}}=\cos [2\pi
(Q_{h}+Q_{k})]+\cos [2\pi (-Q_{h}+Q_{k})]-2$. We have fitted this
expression to the dispersion along the zone boundary in Fig.\
\ref{fig:zone_boundary_dispersion} and the results are $J=0.657\pm
0.002$ meV, $J^{\prime}=0.006\pm 0.003$ meV, compared to
$J=0.673\pm 0.028$ meV, $J^{\prime}=0.012\pm 0.002$ meV obtained
previously\cite{cowley1}. The difference in $J$ arises because the
frequencies at the zone boundary are lower in our measurements due
possibly to small errors in the absolute energy calibration in one
of the two experiments.
%The smaller value for $J^{\prime}$ is
%possibly due to an error in the previous measurements in the
%definition of the exchange constant.

In short the present experiments observe a definite change in
energy along the antiferromagnetic zone boundary. The effect is
small and the data barely produces a reliable estimate of the
effect. Nevertheless, our results do suggest that its most
probable origin is from the next nearest neighbour exchange
constants. We shall however neglect this small effect for most of
the remainder of this paper.

\section{Two-Magnon Continuum Scattering}

\label{TwoMagnon}

\begin{figure}[ht!]
\begin{center}
\includegraphics*[width=8cm,bbllx=11,bblly=80,bburx=576, bbury=820]{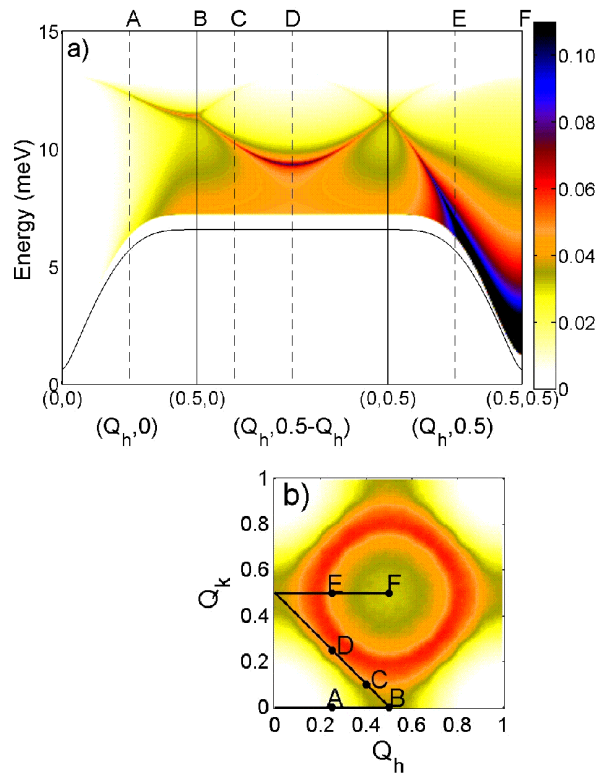}
\end{center}
\caption{(Color online) (a) Two-magnon scattering intensity
$S^{zz}(\bm{Q},\protect\omega)_{\mathrm{inelastic}}$ as a function
of energy and wavevector along symmetry directions in the
Brillouin zone (thick solid curve in (b)). Colour represents
intensity. The solid line shows the one-magnon dispersion relation
$\hbar\omega_{\bm Q}$ from eq.(\ref{eqn:1_S_dispersion}). (b)
Two-magnon intensity as a function of wavevector at constant
energy $\hbar\omega=8.75\pm1.25$ meV. Lineshapes at positions
labelled A-F are shown in Fig.\ \ref{fig:two_magnonx}.
%Constant-$\bm{Q}$ slices are shown in i)$Q_k=0$,
%ii)$Q_h+Q_k=0.5$, iii)$ Q_k=0.5$. The one magnon cross section is
%also plotted as a black line in i)-iii), showing the gap between
%the one and two-magnon excitations. A constant energy slice is
%shown in iv), evaluated between 7.5meV and 10meV. The location of
%the cut lineshapes, A-F, of Fig.~\protect\ref{fig:two_magnonx},
%are also marked.
} \label{fig:two_slices_magnonx}
\end{figure}

To estimate the two-magnon scattering intensity predicted by spin
wave theory we evaluated eq.\ (\ref{eqn:Szz_Rb2MnF4_lowT})
numerically by summation over a grid of finely spaced ${\bm Q}_1$
points over one Brillouin zone, setting $\bm{Q}_{2}=
\bm{Q}-\bm{Q}_{1}$ and replacing the delta function in energy with
an area-normalized narrow Gaussian.
Fig.~\ref{fig:two_slices_magnonx}a) shows an overview plot of the
calculated two-magnon cross-section as a function of energy and
wavevector along symmetry directions the Brillouin zone.
Two-magnon scattering occurs in the form of a continuum at higher
energies above the one-magnon dispersion relation. The small
energy separation $E_{\mathrm{gap}}(\bm{Q})$ between one- and
two-magnon excitations is a consequence of the small uniaxial
anisotropy $\delta_z$ in the Hamiltonian eq.\ (\ref{Hamiltonian}),
which opens a gap in the one-magnon spectrum at the zone centre
$\hbar\omega _{\bm 0}=4JS\sqrt{\delta_z (2+\delta_z)}$, with
two-magnon scattering starting at the higher energy of
$2\times\hbar\omega _{\bm 0}$. Generally the two-magnon intensity
is strongest for low energies and wavevectors near the
antiferromagnetic zone centre, but here is also where the
one-magnon structure factor is largest. The clearest way to
separate a scattering signal from one- and two-magnon processes is
at energies above the one-magnon zone boundary, and Fig.\
\ref{fig:two_slices_magnonx}b) shows a plot of the calculated
two-magnon intensity distribution in the Brillouin zone at an
energy $8.75\pm1.25$ meV, above the one-magnon cut-off.

%bbllx=5,bblly=115,bburx=580, bbury=826

\begin{figure}[ht!]
\begin{center}
\includegraphics*[width=8cm,bbllx=10,bblly=31,bburx=585, bbury=837]{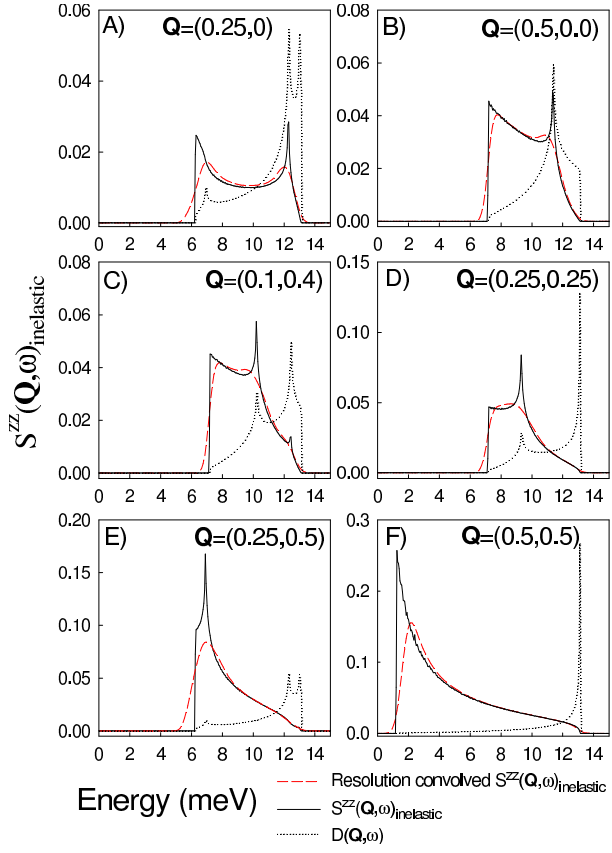}
\end{center}
\caption{Calculated two-magnon scattering lineshapes at fixed
wavevectors, indicated by labels A-F in Fig.\
\ref{fig:two_slices_magnonx}. Solid line is
$S^{zz}(\bm{Q},\omega)_{\mathrm{inelastic}}$ and dashed lines show
the effects of the instrumental resolution. Dotted lines show the
two-magnon density of states, $D(\bm{Q},\omega)$ eq.\
(\ref{eqn:density_of_states}), divided by a factor of 10.
\label{fig:two_magnonx}}
\end{figure}

Calculated lineshapes at representative points in the Brillouin
zone are shown in Fig.\ \ref{fig:two_magnonx}. One noticeable
feature is the appearance of strong singular peaks, which become
more prominent upon increasing the numerical accuracy in
evaluating eq.\ (\ref{eqn:Szz_Rb2MnF4_lowT}). The singularities
are a result of divergencies in the two-magnon density of states
obtained by putting $f=1$ in eq.\ (\ref{eqn:Szz_Rb2MnF4_lowT}),
i.e.
\begin{eqnarray}
D(\bm{Q},\omega)&=&\frac{1}{N}\sum_{\bm{Q}_{1},\bm{Q} _{2}}\delta
(\bm{Q}-\bm{Q}_{1}+\bm{Q}_{2})\delta (\hbar \omega
-\hbar \omega _{\bm{Q}_{1}}-\hbar \omega _{\bm{Q}_{2}})  \notag \\
&=&\frac{1}{N}\sum_{\bm{Q}_{1}}\delta (\hbar \omega -\hbar \omega
_{ \bm{Q} _{1}}-\hbar \omega _{\bm{Q}-\bm{Q}_{1}})
\label{eqn:density_of_states}
\end{eqnarray}
Plots of $D(\bm{Q},\omega)$ are shown by dotted lines in Fig.\
\ref {fig:two_magnonx}. The singularities at the high energy
boundary present in $D(\bm{Q},\omega)$ do not show up in
$S^{zz}(\bm{Q} ,\omega)_{\mathrm{inelastic}}$. This is because the
structure factor of those processes in neutron scattering cancels
as both magnons are on the antiferromagnetic zone boundary contour
where $v_{\bm{Q}_1}=v_{\bm{Q}_2}=0$.

It is interesting to consider whether any of the singularities are
a true feature of the two-magnon neutron scattering or whether
they are a consequence of using non-interacting spin wave theory.
Canali and Wallin \cite{Canali2} have included first order spin
wave interactions in calculating $S^{zz}(\bm{Q},\omega)$ and their
results show that the singularity peaks remain. However, any
treatment of interactions within spin wave theory is perturbative
and so it is possible higher order interactions may be still be
important when considering the singularities. Experimentally,
resolution effects would make the singularities very difficult to
observe as shown in Fig.\ \ref{fig:two_magnonx} where the
two-magnon lineshapes are convoluted with the resolution in our
experiment and the results (dashed lines) show that any
singularity would probably not be visible in our measurements.

\begin{figure}[th]
\begin{center}
\includegraphics*[width=8cm,bbllx=53,bblly=41,bburx=558, bbury=780]{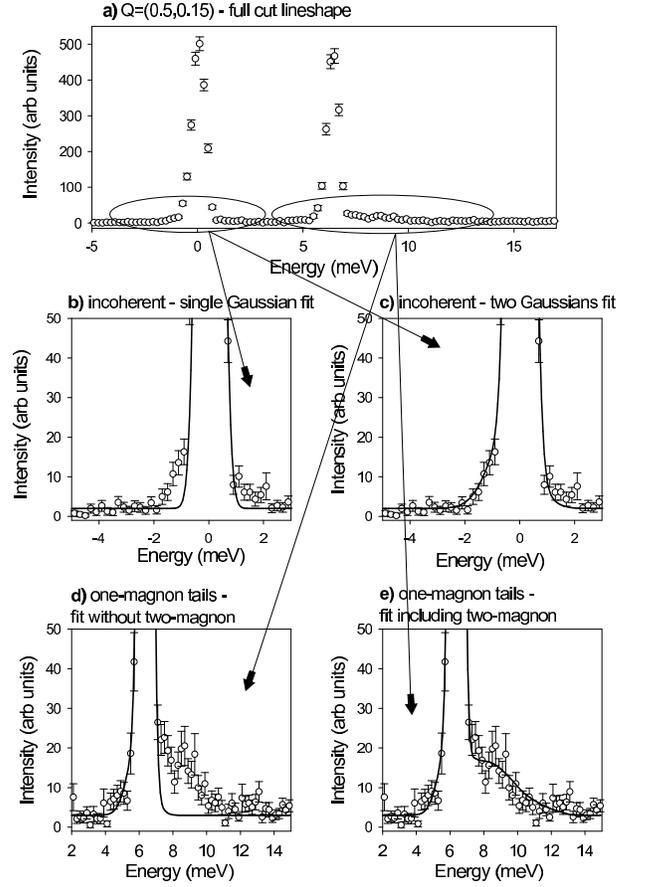}
\end{center}
\caption{Stages of the fitting procedure. a) Energy scan at
(0.5,0.15). b)-c) Low-energy asymmetric lineshape of the
incoherent scattering is well parameterized by a two Gaussian
lineshape. d) Fitting the same resolution profile to the
one-magnon peak centred at 6.40 meV, the extra scattering on the
high-energy tail can not be accounted for by resolution effects.
e) Fits to the combined one- and two-magnon scattering eq.\
(\ref{eqn:scattering_x_section}) including resolution effects.}
\label{fig:two_magnon_fits}
\end{figure}

To test for the presence of two-magnon scattering in the data, we
extracted energy cuts at fixed wavevectors ($Q_h,Q_k$). Because of
the inherently weak intensity of the two-magnon cross-section we
chose a wavevector averaging range $0.1 \times 0.1$ to have enough
data pixels for quantitative analysis. A typical scan near the
antiferromagnetic zone boundary at (0.5,0.15) is shown in Fig.\
\ref{fig:two_magnon_fits}a). The inelastic scattering is dominated
by a sharp, one-magnon peak centred at 6.40 $\pm$ 0.02 meV, and
additional much weaker scattering is observed in the form of a
high energy continuum tail extending to at least 9 meV [see Fig.\
\ref{fig:two_magnon_fits}d)], much higher that the one-magnon zone
boundary energy.

%The energy-integrated intensity of this high-energy tail is \% of the
%integrated intensity of the one-magnon peak, which is consistent
%with the intensity ratio expected for two- and one-magnon
%scattering according to eqs.\ (\ref{eqn:2m_scattering_x_section}).

Our approach is to fit the data to a lineshape containing both one
and two-magnon components, with their relative intensity fixed by
theory, eqs.\ (\ref{eqn:one_magnon_corellations}),
(\ref{eqn:Szz_Rb2MnF4_lowT}) and (\ref{eqn:scattering_x_section}).
The effects of resolution broadening are also included as
discussed below. The effects of the finite cut size were included
by averaging the predicted intensity over the finite wavevector
size $\Delta Q_h \times \Delta Q_k$ of the cut. The resulting
profile was then convolved with the energy resolution of the
spectrometer. This was determined from the observed lineshape of
the quasi-elastic peak in Fig.\ \ref{fig:two_magnon_fits}b). This
showed a slightly asymmetric tail at lower energies (due to the
asymmetric neutron pulse shape) and the whole profile could be
well parameterized by a sum of two Gaussian peaks, one off-centred
on the low-energy side, see Fig.\ \ref{fig:two_magnon_fits}c).
Such a weakly-asymmetric lineshape also provided a good
description of the observed one-magnon peak lineshape as shown in
Fig.\ \ref{fig:two_magnon_fits}d). The relative positions,
intensities and widths of those two Gaussians are fixed while the
fitted parameters are the overall width of the one-magnon peak (to
parameterize the variation of the energy resolution with energy
transfer) and an overall scale factor. Fig.\
\ref{fig:two_magnon_fits}e) shows that the whole observed
scattering lineshape including the high-energy tail can be well
described when the two-magnon cross-section is included.

%The results of some two-magnon lineshapes after the above
%resolution correction are shown in Fig.\ \ref{fig:two_magnonx}
%(dashed line): the main effects of resolution broadening is a
%smearing out the singularity peaks, with the overall lineshape
%left largely unchanged.

\begin{figure}[ht!]
\begin{center}
\includegraphics*[width=8cm,bbllx=95,bblly=280,bburx=564, bbury=737]{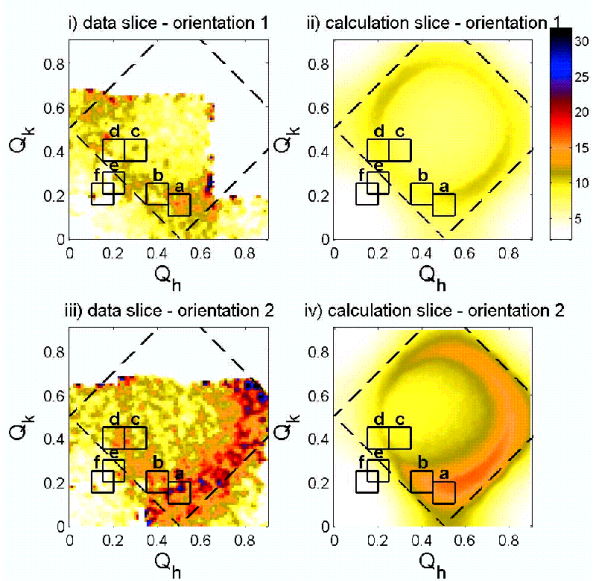}
\end{center}
\caption{(Color online) Comparison between data and predicted
two-magnon scattering at an energy $\hbar\omega=8.75\pm1.25$ meV,
much higher that the one-magnon zone boundary energy. i) and iii)
show data in two different experimental configurations and ii)and
iv) show the simulated two-magnon scattering intensity including
the polarization factor, magnetic form factor, an overall scale
factor and a flat background to compare with data. Boxes labelled
a)-f) show location of energy scans plotted in Fig.\
\ref{fig:two_magnon_cuts}. The dashed square box is the
antiferromagnetic zone boundary.} \label{fig:all_slices}
\end{figure}

Fig.\ \ref{fig:all_slices} shows the overall comparison between
the observed intensity in the high-energy tails and that expected
from two-magnon scattering. The plot corresponds to intensity
averaged over the energy range 7.5 to 10 meV, much higher than
that of the one-magnon zone boundary, where only two-magnon
processes are expected to contribute. There is good overall
agreement between data and calculations, which include the
longitudinal polarization factor $p_z$ expected for two-magnon
processes eq\ (\ref{eqn:scattering_x_section}).

%A qualitative comparison can be made between the two-magnon cross
%section and the experimental data, by comparing some constant
%energy slices, as shown in Fig.~\ref{fig:all_slices}. These slices
%are taken between 7.5meV and 10meV, which are energies higher than
%the one-magnon excitations and in the region were the two-magnon
%excitations have a strong intensity. The colour plots show a good
%qualitative similarity.

%two_magnon_slices_gammaii_colorbar3.eps

\begin{figure*}[t!]
\begin{center}
\includegraphics*[width=17.5cm,bbllx=78,bblly=472,bburx=536, bbury=789]{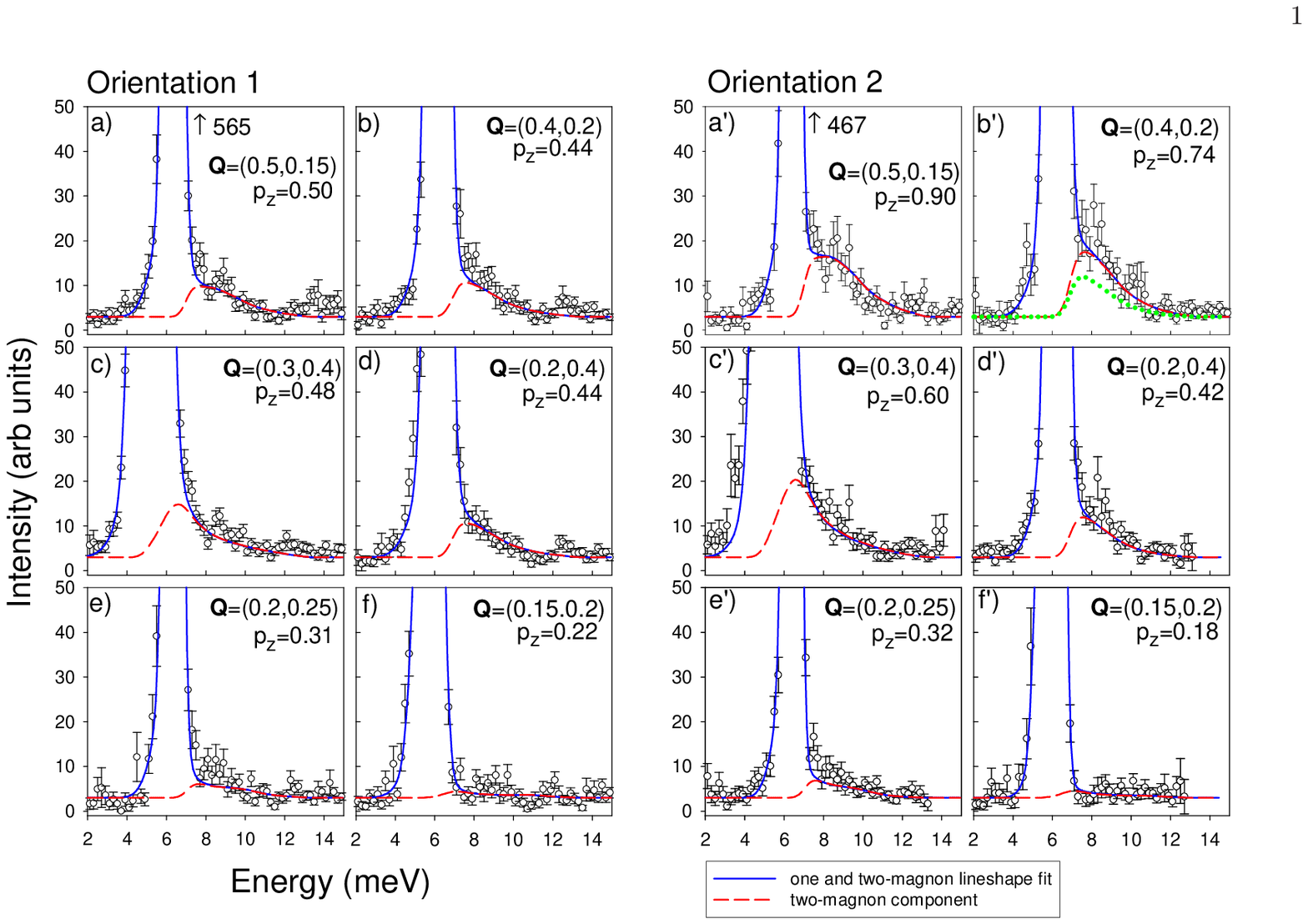}
\end{center}
\caption{(Color online) Energy scans analysed with a combined one
and two-magnon cross-section eq.\ (\ref{eqn:scattering_x_section})
(solid lines). Dashed lines shows the two-magnon component, with a
common overall scale factor for all scans, as used in the
simulation in Fig.\ \ref{fig:all_slices}. Scan labels a-f) (with
$\prime$ for data in orientation 2) refer to locations in the
Brillouin zone indicated in Fig.\ \ref{fig:all_slices}. $p_z$ is
the longitudinal polarization factor in the middle of the
two-magnon continuum region at 8.75 meV. Dotted line in
b$^{\prime}$) illustrates the type of disagreement obtained if one
uses the two-magnon intensity observed at the equivalent position
d$^{\prime}$) without adjusting for the change in polarization
factor. \label{fig:two_magnon_cuts}}
\end{figure*}

A number of representative energy scans extracted from the data
near the antiferromagnetic zone boundary where the two-magnon
contribution can most easily be singled out are shown in Fig.\
\ref{fig:two_magnon_cuts}. Solid lines show the results of the
fitting procedure described above to the combined one-and
two-magnon scattering lineshapes eq.\
(\ref{eqn:scattering_x_section}) and good agreement is observed
throughout. In those fits the overall scale factor was allowed to
vary for each scan, however very similar fits are obtained if this
scale factor is fixed to a common average value for all scans and
for clarity we only show the two-magnon component determined in
this way (dashed lines), practically indistinguishable  from the
results of the free fits. As an independent consistency check we
have converted the neutron intensities into absolute units of barn
meV$^{-1}$ sr$^{-1}$ per spin normalizing by the sample mass and
by intensities measured with a vanadium standard; the calculated
two-magnon neutron scattering intensities in absolute units are
very close (within $10\%$) to the results of the fits, this
agreement giving further support to the identification of the
continuum intensities with two-magnon scattering.

The scans shown in Fig.\ \ref{fig:two_magnon_cuts} include points
where the intensity from two-magnon scattering is predicted to be
low, as well points where it is predicted to be high, this
modulation of intensity being mainly due to the polarization
factor $p_z$ in eq.\ (\ref{eqn:scattering_x_section}). For example
scans in Fig.\ \ref{fig:two_magnon_cuts}e)-f) have rather weak
high-energy tails and cannot be taken in isolation to provide
evidence of two-magnon scattering, but are significant when taken
in the context of all the scans shown. The scans in Fig.\
\ref{fig:two_magnon_cuts}a$^{\prime}$)-b$^{\prime}$) show the
strongest high-energy signal, since they correspond to the largest
longitudinal polarization factor appropriate for two-magnon
scattering (and {\em lowest} polarization factor for transverse
one-magnon scattering). Another illustration of the polarization
effect is provided by comparing the data in Figs.\
\ref{fig:two_magnon_cuts}b$^{\prime}$) and d$^{\prime}$) collected
at equivalent positions in the Brillouin zone: using the same
two-magnon intensity in both scans without adjusting for the
change in polarization factor results in a large disagreement
(dotted line in Fig.\ \ref{fig:two_magnon_cuts}b$^{\prime}$) with
the observed continuum scattering intensity. From this analysis we
conclude that the observed continuum scattering at high energies
is consistent both in magnitude and polarization with scattering
expected from two-magnon processes, neglecting any interactions
between the magnons.

\section{Conclusions}
\label{Conclusions}

We have conducted a detailed investigation of the low temperature
dynamical properties of the square-lattice spin-5/2 Heisenberg
antiferromagnet Rb$_2$MnF$_4$. The spin wave dispersion was
measured and a small variation in energy along the
antiferromagnetic zone boundary with found. The energy change
along the zone boundary was too large to be explained by quantum
corrections to linear spin wave theory, and could be the result of
weak next-nearest neighbour exchange interactions
($J^{\prime}/J=1\pm0.5$\%). Furthermore, a low intensity signal
was observed around the high energy tail of the one-magnon peaks.
The lineshape and intensity variation of this signal provides good
evidence that it is the result of scattering by pairs of
non-interacting spin waves (two-magnon scattering). We conclude
that although spin wave interactions are important in describing
the shape of the two-magnon Raman scattering, they are much less
important for two-magnon neutron scattering.

This research was supported by UK EPSRC (TH, RC, RAC, DAT) and by
US NSF through grant number DMR-0134377 (RLL). We are also
grateful to the staff at ISIS for their help and support and to
Bella Lake and Robert Birgeneau for useful discussions.

{
\bibliographystyle{plain}
\bibliography{D:/MapsRb2MnF4/Two_Magnon_Paper/TBib}
}

\end{document}